\RequirePackage{fix-cm}
\documentclass{svjour3}                     % onecolumn (standard format)
\smartqed  % flush right qed marks, e.g. at end of proof
\usepackage{graphicx}
\usepackage{amsmath}
\usepackage{latexsym}
\newsavebox{\astrutbox}
\sbox{\astrutbox}{\rule[-5pt]{0pt}{20pt}}

\def\der#1#2{{\partial #1\over \partial #2}}

\def\be{\begin{equation}}
\def\ee{\end{equation}}
\def\bea{\begin{eqnarray}}
\def\eea{\end{eqnarray}}
\def\bse{\begin{subequations}}
\def\ese{\end{subequations}}
\def\bsea{\begin{subeqnarray}}
\def\esea{\end{subeqnarray}}
\def\({\left (}
\def\){\right )}
\def\[{\left [}
\def\]{\right ]}
\def\<{\left <}
\def\>{\right >}

\begin{document}

\title{On entropy production in the Madelung fluid and the role of Bohm's potential in classical diffusion}
\titlerunning{}

\author{Eyal Heifetz, Roumen Tsekov, Eliahu Cohen and Zohar Nussinov}


\institute{Eyal Heifetz \at Department of Geosciences, Tel-Aviv University, Tel-Aviv, Israel\\
  {\email eyalh@post.tau.ac.il}
	\and Roumen Tsekov \at Department of Physical Chemistry, University of Sofia, 1164 Sofia, Bulgaria\\
  {\email Tsekov@chem.uni-sofia.bg}\
	\and Eliahu Cohen \at H.H. Wills Physics Laboratory, University of Bristol, Tyndall Avenue, Bristol, BS8 1TL, U.K\\
  {\email eliahu.cohen@bristol.ac.uk}\	
	\and  Zohar Nussinov
	\at Department of Physics, Washington University, St. Louis, MO 63160, U.S.A. \\
	{\email zohar@wuphys.wustl.edu}\\
	 }


\date{\today}

\maketitle

\begin{abstract}

The Madelung equations map the non-relativistic time-dependent\\ Schr\"{o}dinger equation into hydrodynamic equations of a virtual fluid.
Here we show that an increase of the Boltzmann entropy of this Madelung fluid is proportional to the expectation value of its velocity divergence. Hence, entropy growth is accompanied by expansion resulting from the ability of the Madelung fluid to be compressible. The compressibility itself reflects superposition of solutions of the Schr\"{o}dinger equation. Thus, in unitary processes where the Madelung fluid expands and then shrinks, the Boltzmann entropy may, correspondingly, grow and then decrease. The notion of entropy growth due to expansion is common in diffusive processes, however in the latter the process is irreversible. Much unlike the Boltzmann entropy, the von Neumann entropy, does not vary with time. To elucidate the physical underpinning of the Boltzmann entropy,
we examine several specific examples. We demonstrate that, for classical diffusive processes, the ``force'' accelerating diffusion has the form of the positive gradient of the quantum Bohm potential.
In the Madelung fluid, the advective and the diffusive velocities correspond respectively to the the real and imaginary parts of the complex momentum. We find that the diffusion coefficient provides a lower bound of Heisenberg uncertainty type product between the gas mean free path and the Brownian momentum.

\end{abstract}

\section{Introduction}
%\label{intro}

The Madelung equations \cite{Mad} transform the non-relativistic time-dependent\\ Schr\"{o}dinger equation into hydrodynamical equations of an Eulerian fluid
\cite{Tak,Sch,Son,HC}. The aim of this study is to explore the entropy properties of the Madelung fluid and compare it with the ones of classical Eulerian fluids.

The hydrodynamical transformation is obtained when considering the Schr\"{o}dinger equation,
\be
\label{sc}
i\hbar\der{\Psi}{t} = \hat{H}\Psi = \({\hat{p}^2 \over 2m} +U\)\Psi =
\(-{\hbar^2\over 2m}\nabla^2 +U\)\Psi,
\ee
for a continuous wave function $\Psi({\bf r},t) = \sqrt{\rho}({\bf r},t)e^{iS({\bf r},t)/\hbar}$ (so that $\rho = \Psi^*\Psi$) of a particle with mass $m$, in the presence of an external potential $U({\bf r},t)$. Using the de Broglie guiding equation, ${\bf u}_a = \nabla{\tilde S}$ (where the tilde superscripts represent hereinafter quantities per unit mass $m$), the real part of (1) becomes the continuity equation
\be
\der{\rho}{t} = - \nabla\cdot{(\rho {\bf u}_a)}.
\ee
The imaginary part of the  Schr\"{o}dinger equation becomes the Eulerian fluid momentum equation,
\be
{D_a\over Dt}{\bf u}_a = -\nabla \tilde{Q}(\rho) -\nabla{\tilde U}.
\ee
Here, ${D_a\over Dt} \equiv \der{}{t} + {\bf u}_a\cdot\nabla$, is the material (advective) time derivative of a fluid element along its trajectory and
$\tilde{Q} = - {\hbar^2 \over 2m^2}{\nabla^2 \sqrt{\rho} \over \sqrt{\rho}}$ is the Bohm potential per unit mass \cite{Bohm}.

In Ref. \cite{HC}, it was suggested that the conservation of the domain integrated  energy in (1) implies that the Madelung fluid is adiabatic. The fluid conserves the sum of the domain integrated kinetic, potential and internal energy, where the latter is given by the Fisher information. The domain averaged adiabaticy is in agreement with the conservation of the von Neumann entropy. The von Neumann entropy, i.e., the trace $Ent_{VN}=  - k_{B}~Tr[\hat{\rho} \ln \hat{\rho}]$, where $k_{B}$ is the Boltzmann constant and $\rho$ the density matrix of a closed system with a general (possibly time dependent) Hamiltonian, cannot (by virtue of
unitary time evolution) change with time. One may, nevertheless, devise other illuminating entropy functionals (e.g., the ``diagonal entropy'' of \cite{polkovnikov})
that, even for closed systems, transparently adhere to standard thermodynamic relations (including the second law of thermodynamics).
In the current work, our focus is on quantities associated with the Madelung fluid.
Domain averaged adiabaticy of the Madelung fluid does not imply that entropy is materially conserved by a ``fluid parcel''. This viable non-conservation differs from that of the thermodynamic entropy, $\tilde{\eta} = \ln{(T^{Cv}/\rho^{Nk_{B}})}$, where $C_v$ is the specific heat of an $N$ particle adiabatic ideal gas, and ${D_a\over Dt}\tilde{\eta}=0$  (see, e.g., \cite{Vallis}).
We will find that the classical Madelung fluid dynamics indeed motivates the introduction of a ``Boltzmann entropy of the Madelung fluid'' ($Ent_B$) and illustrates that the dynamics is intuitively appealing.
Specifically, for the standard density $\rho \equiv \Psi^*\Psi$ in, e.g., real space, we set
\be
Ent_B \equiv -k_{B} \int \rho \ln \rho dV \equiv k_{B} \int s dV.
\ee
For Joule free expansion of an ideal gas into a vacuum the temperature remains constant, the standard increase of the thermodynamic entropy is captured by the change in $Ent_B$. The non-trivial temporal changes of the Boltzmann entropy of (4) are, generally, very different from those of the von Neumann entropy. The Boltzmann entropy is intimately linked to the Shannon information \cite{Wehrl}.

In Section 2, we explicitly discuss the conservation of the von Neumann entropy for a diagonal density matrix $\hat{\rho}$. We underscore that by virtue of
the invariance of the trace defining $S_{VN}$ under unitary time evolution, the von Neumann entropy is time independent for any initial density matrix (whether diagonal or not).
The new form of $S_{VN}$ that we derive will lucidly relate it to the Madelung fluid density $\rho$. In Section 3, we will examine the Boltzmann entropy production in the Madelung fluid via expansion by considering examples of both reversible and irreversible processes.

The continuity equation (2) represents advective rather than diffusive dynamics, where the negative value of the Bohm potential gradient acts as a force in the Madelung fluid (3). Nevertheless, in Section 4 we show that for classical diffusion the positive value of the Bohm potential gradient acts as an effective force. Besides the realization of the quantum Bohm potential in a classical phenomenon this implies that the Bohm potential represents diffusive processes in the Madelung fluid, as discussed in Section 5. Summary and conclusions appear in Section 6.

\section{The von Neumann entropy conservation in the Madelung fluid}

In the sections that follow we will largely focus on the Boltzmann entropy associated with the Madelung fluid. Before doing so, we will now briefly derive the von Neumann entropy $S_{VN}$ for this fluid. For a continuous wave function $|\psi\rangle=\int\psi({\bf r})|{\bf r}\rangle dV$, the density matrix takes the integral form
\be
\hat{\rho}=|\psi\rangle\langle\psi|=\int\int  \psi({\bf r})\psi^*({\bf r}')|{\bf r}\rangle\langle {\bf r}'| dV dV'.
\ee
In the Appendix it is shown explicitly that, for a diagonal $\hat{\rho}$, the von Neumann entropy then becomes
\be \label{vne}
Ent_{VN}=-Tr[\hat{\rho} \ln\hat{\rho}]= -\int\int \psi^*({\bf r})\psi({\bf r}')\ln[\psi^*({\bf r})\psi({\bf r'})]dV dV'.
\ee
As it must, $\der{}{t}Ent_{VN} = 0$. A direct calculation (see the Appendix) reveals that
\begin{equation}
\begin{array}{lcl}
\label{evn}
Ent_{VN}= -\int\int dV dV' \\
\sqrt{\rho({\bf r})\rho({\bf r}')}\[\ln{\sqrt{\rho({\bf r})\rho({\bf r}')}}\cos\({S({\bf r}) -S({\bf r}')\over\hbar}\)
+\({S({\bf r}) -S({\bf r}')\over\hbar}\)\sin\({S({\bf r}) -S({\bf r}')\over\hbar}\)  \].
\end{array}
\end{equation}

Eq. (\ref{evn}) illustrates that the conserved von Neumann entropy is fundamentally different from the Boltzmann entropy of Eq. (4). $S_{VN}$ is a measure of the spatial correlation between different properties within the fluid. Furthermore, it requires explicit information on the velocity potential $S$ and not just on the fluid density $\rho$. To the best of our knowledge it is not related to any classical measure of fluid entropy, thus its conservation is applied uniquely to the Madelung quantum fluid.

\section{The Boltzmann Entropy production and compressibility effects}

Within the definition of (4), the continuity equation (2) yields
\be
\der{s}{t} + \nabla\cdot{(s {\bf u}_a)} = \rho\nabla\cdot{\bf u}_a,
\ee
and if all fluxes vanish at the domain boundaries, we immediately obtain that
\be
\der{}{t}Ent_B = k_{B}\int \rho\nabla\cdot{\bf u}_a dV = k_{B}\< {\nabla\cdot{\bf u}_a} \> = k_{B} \< {\nabla^2{\tilde S}}\>.
\ee
Hence, the total entropy production is equal to the expectation value of the divergence, i.e., the Boltzmann entropy grows through expansion of the fluid and may decay through compression. For the Madelung fluid to be compressible, the quantum action $S$ must
have a ``source'' in the sense that it has to satisfy some Poisson equation in the form of $\nabla^2{ S} \neq 0$.
From the wave function perspective, compressibility results from superposition. For a single plane wave solution of the form
$\Psi = \sqrt{\rho} e^{i({\bf k}\cdot{\bf r} - \omega t)}= \sqrt{\rho} e^{iS / \hbar}$,  ${\bf u}_a =  \nabla{\tilde S} = {\hbar\over m}{\bf k}$. Thus, the advective velocity is simply proportional to the wavenumber ${\bf k}$ and $\nabla\cdot {\bf u}_a = 0$. However, when two plane waves or more interfere,  $\nabla\cdot {\bf u}_a \neq 0$ in general.

In the next two simple examples we show how a superposition of plane waves triggers entropy growth. We will furthermore see how compressibility may lead to a reversible process. In the two cases we consider, the dynamics of a 1D Gaussian density solution has been derived by \cite{Tsekov92},
\be
\label{Gauss}
\rho(x,t) =  {1\over \sigma(t) \sqrt{2\pi}}e^{-{x^2 \over 2\sigma^2(t)}}\, .
\ee
Substituting (10) in the continuity equation (2) yields
\be
u_a(x,t) = x\der{\ln\sigma}{t}
\hspace{0.25cm} \Longrightarrow \hspace{0.25cm} \nabla\cdot{\bf u}_a = \der{\ln\sigma}{t}.
\ee

(i) First we consider the case of a free particle. Substituting (10) and (11) in the 1D version of (3) for $\tilde{U} = 0$ yields
\be
\sigma\der{^2\sigma}{t^2} = \({\hbar \over 2m\sigma}\)^2 \hspace{0.25cm} \Longrightarrow  \hspace{0.25cm}
\sigma^2 = \sigma_0^2 + \({\hbar t\over 2m\sigma_0}\)^2\, ,
\ee
with the Boltzmann entropy
\be
Ent_B = k_{B}\ln{(\sigma \sqrt{2\pi e})} ={Ent_B}_0 +{k_B\over 2}\ln\[1+\({\hbar t\over 2m\sigma_0^2}\)^2\]\, ,
\ee
so that
\be
{1\over k_{B}}\der{Ent_B}{t} = \nabla\cdot{\bf u}_a = \der{\ln\sigma}{t} = {t \over \({2m\sigma_0^2\over \hbar}\)^2 +t^2} > 0.
\ee

These results illustrate how the superposition of plane waves in the Gaussian wave packet of Eq. (\ref{Gauss}) influences the compressibility of the Madelung fluid and how it gives rise to an increase of entropy. For a free particle, a 1D plane wave solution of (1) has the form of $\Psi = A(k)e^{ik[x-({\hbar\over 2m})kt]}$, hence the dynamic Gaussian solution of (10) is a continuous superposition of plane waves whose amplitude $A(k) \propto e^{-{(\sigma_0 k)^2}}$, as can be verified from the Fourier transform of (10) at time $t=0$.

(ii) As a second example we solve (10) in the presence of the harmonic potential  $U = {m \over 2}(\omega_0 x)^2$ to obtain from (3):
\be
\sigma\der{^2\sigma}{t^2} = \omega_0^2(\sigma^2_0 - \sigma^2), \hspace{0.5cm} \sigma_0^2 = {\hbar\over 2 m \omega_0},
\ee
which can be solved numerically. The well known stationary ground state solution (e.g. \cite{LL}) in which $\sigma = \sigma_0$, is a special case of (15) where
$\rho_0 =  {1\over \sigma_0 \sqrt{2\pi}}e^{-{x^2 \over 2\sigma_0^2}}$, and the Boltzmann entropy is constant.
\be
{Ent_B}_0 = -k_{B} \int \rho_0\ln\rho_0 dx = k_{B}\ln{(\sigma_0 \sqrt{2\pi e})}.
\ee
As pointed out by \cite{HC}, and is evident from (11), the only possible solution for the velocity in this case is ${u}_a =0$, hence the Madelung fluid is obviously incompressible in the ground state. Consider, however, a small deviation from the ground state:
$\sigma = \sigma_0 + \epsilon(t)$, where $|\epsilon|/\sigma_0 << 1$. Eq. (15) then yields $\der{^2 \epsilon}{t^2} = -2\omega_0^2\epsilon +O(\epsilon^2)$, so that
$\epsilon(t) = {\epsilon}_0\cos(\sqrt{2}\omega_0t)$ for $O(\epsilon)$. Hence,
\be
 Ent_B = {Ent_B}_0 + k_{B}({\epsilon \over \sigma_0}), \hspace{0.25cm}
{1\over k_{B}}\der{Ent_B}{t} = \nabla\cdot{\bf u}_a ={1\over \sigma_0}\der{\epsilon}{t} = -{\sqrt{2}\omega_0\over \sigma_0}\sin(\sqrt{2}\omega_0t)\, ,
\ee
implying a reversible sinusoidal variation of the Boltzmann entropy.

For completeness we note that the action $\tilde{S}$, associated with the wave function of (10), can be found explicitly (up to some constant) using $u_a = x\der{}{t}{\ln\sigma} = \der{}{x}\tilde{S}$ so that
$\tilde{S}= {x^2 \over 2}\der{\ln\sigma}{t} +f(t)$. The time-dependent function $f(t)$, must satisfy the Hamilton-Jacobi equation (or the time-dependent Bernoulli
equation in the fluid dynamics language \cite{HC}), which is the imaginary part of (1) from which (3) is derived. For the 1D version this equation becomes
\be
\der{\tilde S}{t} + {1\over 2}\(\der{\tilde S}{x}\)^2 +\tilde{Q} + \tilde{U} = 0,
\ee
yielding $f(t) = - ({\hbar\over 2 m })^2\int_{t_0}^t{dt\over \sigma^2}$.

%We briefly consider the superposition $\Psi = c_{1} \Psi_1 + c_{2} \Psi_{2}$ of two non-overlapping wavefunctions. As a consequence of entropy concavity (see \cite{Wehrl}, for instance), i.e., $s(\rho)\ge\lambda_1 s(\rho_1)+\lambda_2 s(\rho_2)$, where $\rho=\lambda_1 \rho_1 + \lambda_2 \rho_2$, with non-negative normalized weights (i.e., $\lambda_1= |c_1|^{2},\lambda_2=|c_{2}|^{2} \ge 0$ and $\lambda_1+\lambda_2=1$), we observe that such a superposition may lead to an entropy increase.
%In the Madelung framework such interference yields compressibility.

%In the case of a single plane wave of wave number ${\bf k}$, ${\bf u}_a =  \nabla{\tilde S} = {\hbar\over m}{\bf k}$ so that $\nabla\cdot {\bf u}_a = 0$. However, when two plane waves or more interfere,  $\nabla\cdot {\bf u}_a \neq 0$ in general, so that the entropy may grow according to (6).

\section{The role of the Bohm potential in classical diffusion}

%Expansion of fluids increases their entropy as it increases the amount of its microstates.
%In the Madelung fluid the expansion is due to the hydrodynamic velocity and therefore processes can be reversible in principle. This stands in contrast with irreversible diffusive processes such as Joule expansion. Here we wish to compare between the two processes more closely.

We wish to compare the entropy dynamics of the Madelung fluid to the irreversible processes obtained in classical diffusion. We consider the standard case where the diffusive velocity
${\bf u}_{d}$ satisfies Fick's first law:
\be
{\bf u}_{d} = -D\nabla\ln\rho,
\ee
where $D$ is the diffusion coefficient (assumed constant for simplicity).
Changes in density result then from diffusive fluxes (rather than advective fluxes in the hydrodynamic continuity equation of (2)) as stated by Fick's second law:
\be
\der{\rho}{t} = -\nabla\cdot{(\rho {\bf u}_{d})} = D \nabla^2 \rho\, .
\ee
Equation (8) still holds when ${\bf u}_{a}$ is replaced by ${\bf u}_{d}$, but its domain integration yields now:
\be
\der{}{t}(Ent_B)_D = k_{B}\int \rho\nabla\cdot{\bf u}_d dV = k_{B}\< {\nabla\cdot{\bf u}_d} \> = k_{B} D   \<(\nabla\ln\rho)^2\> \ge 0,
\ee
where $Fi =  \int {(\nabla\rho)^2\over \rho} dV =  \<(\nabla\ln\rho)^2\> $ is the Fisher information and $(Ent_B)_D$ represents the Boltzmann entropy undergoing a diffusion process. Hence, this well known relation (e.g. \cite{Wehrl}, and other information theory contexts) suggests that entropy increases in diffusive processes through expansion, as in the hydrodynamic case, but in contrast with the latter, the entropy increases irreversibly with time as long as density gradients exist (in agreement with the second law of thermodynamics).

%One may ask then if classical diffusive processes have any relation with the dynamics of the quantum Madelung fluid (for clarity we note that [Roumen please add refs...] treat diffusion in the Madelung fluid by adding an additional  diffusive term to (3) and [Roumen please add refs...] treat the Schr\"{o}dinger equation as a diffusion one by using the Wick transformation $t \rightarrow -it$, however here we refer to equation set (1-3) as is). In the next example we show the surprising appearance of the quantum Bohm potential in classical diffusion, where in the next section we consider classical diffusive processes in the quantum Madelung fluid.

Equation (20) can be transformed into a momentum like equation when defining a ``material diffusive derivative'' as ${D_d\over Dt} \equiv \der{}{t} + {\bf u}_d\cdot\nabla$. Then it is straightforward to show that (20) can be translated to
\be
{D_d\over Dt}{\bf u}_d = \nabla \tilde{Q}_d,
\ee
where $\tilde{Q}_d = - 2D^2{\nabla^2 \sqrt{\rho} \over \sqrt{\rho}}$ may be denoted as a diffusive Bohm potential.
Hence, the positive gradient (as opposed to the negative sign in (3)) of the diffusive Bohm potential acts as a ``force'' to accelerate the diffusion. This appearance of the quantum Bohm potential in a classical process is intriguing.

The role of the gradient of the Bohm potential as a diffusive force becomes more transparent when returning to the dynamic 1D Gaussian example. The well known solution to the 1D version of (20) (e.g., \cite{KC}) is given by
\be
\rho(x,t) =  {1\over \sigma(t) \sqrt{2\pi}}e^{-{x^2 \over 2\sigma^2(t)}}, \hspace{0.25cm} \sigma^2 = 2Dt, \hspace{0.25cm}
u_d(x,t) = -D\der{\ln\rho}{x} = x\der{\ln\sigma}{t} = {x\over 2t}.
\ee
Hence
\be
{1\over k_{B}}\der{}{t}(Ent_B)_D = \nabla\cdot{\bf u}_d = {1\over 2t},
\ee
(note that for large $t$ both (14) and (24) experience asymptotic entropy growth which is proportional to $t^{-1}$).
Since $u_d  = {dx \over dt} = {x/ (2t)}$ a fluid element located at $x_0$ at time $t_0$ will be drifted  at time $t$ to $x= \({x_0 \over \sqrt{t_0}}\)\sqrt{t}$
(random walk). Therefore, $u_d =  {x/ (2t)} = \({x_0 \over 2\sqrt{t_0}}\)/ \sqrt{t}$, and the acceleration of the fluid element is
\be
{d \over dt}u_d = - \(x_0 \sqrt{t}\over \sqrt{t_0}\){1\over 4t^2} = -{x \over 4t^2},
\ee
where ${d\over dt}u_d  \equiv {D_d\over Dt}{u}_d \equiv \der{u_d}{t} + {u}_d\der{{u}_d}{x}$.
Thus, in this example the diffusion rate is being decelerated by the gradient of the diffusive Bohm potential
\be
{D_d\over Dt}{u}_d = \der{\tilde{Q}_d}{x} = -2D^2\der{}{x}\({x\over \sigma^2}\)^2 = -{x \over 4t^2}.
\ee

\section{Representation of diffusion in the Madelung fluid}

The authors of \cite{Esp,Nu,HC} defined the complex velocity derived from the momentum operator $-i\hbar\nabla\Psi$, as
\be
{\bf v} =[-i{\hbar\over m}\nabla\ln\Psi] = {\bf v}_r +i{\bf v}_i,
\ee
so that
${\bf v}_r = {\bf u}_a = \nabla {\tilde S}$ is the advective velocity and
${\bf v}_i = -{\hbar \over 2m}\nabla(\ln\rho)$. One may interpret ${\bf v}_i = {\bf u}_{d}$, as in (19), suggesting the relation between the Planck constant and the diffusion coefficient to be ${\hbar \over 2} = mD$, so that ${Q}_d$ becomes identical to the quantum Bohm potential.
For the simplest case of the Einstein relations in an ideal gas, $mD = ({\bf l}\cdot\bar{\bf p})/3$, where $|{\bf l}|$ is the molecular mean free path and $|\bar{\bf p}|$ is the magnitude averaged thermal (random walk) molecular momentum in-between collisions. Isotropy results in the relation ${\hbar \over 2} = l_x p_x$, and indeed $l_x$ and $p_x$ are the basic scales obtained from statistical mechanics for deriving the kinetic theory of gases. In other words, one cannot resolve the ideal gas dynamics within length scales smaller than $l_x$ or for momenta smaller than $p_x$. As pointed out by \cite{Garb1,Garb2}, it is intriguing that these two fundamental scales form the canonical variables which set the exact limiting case of the Heisenberg uncertainty principle.

Incorporating (2) with (19) we obtain the Fokker-Planck equation
\be
\der{\rho}{t} + \nabla\cdot{[\rho({\bf u}_a-{\bf u}_{d})]} = \({\hbar \over 2m}\) \nabla^2 \rho,
\ee
which corresponds to the entropy equation
\be
\der{s}{t} + \nabla\cdot{[(s-\rho){\bf u}_{a}]} =  \({\hbar \over 2m}\)^{-1}{\rho}{\bf u}_a\cdot{\bf u}_{d},
\ee
yielding
\be
\der{}{t}Ent_B = \({\hbar \over 2m}\)^{-1}{k_{B}} \<{\bf u}_a\cdot{\bf u}_{d} \>.
\ee
Thus stating that positive entropy production in the Madelung fluid occurs when the advective and the diffusive velocities are positively correlated within the fluid domain.

%Finally we point out that (3) and (19) yield
%\be
%{D_a\over Dt}{\bf u}_a + {D_d\over Dt}{\bf u}_d = -\nabla{\tilde U},
%\ee
%suggesting that the force imposed by an external potential generates both the advective and the diffusive accelerations, where in its absence the two accelerations are equal with opposite signs.

\section{Summary and Discussion}

Having an intuitive classical interpretation, the Madelung formulation is natural for discussing quantum dynamics and thermodynamics. In Ref. \cite{HC}, two of us highlighted the role of compressibility in linking hydrodynamical and thermodynamical processes in the Madelung fluid.

In the current work, when examining entropy production in this framework, we find that it is proportional to the expectation value of $\nabla\cdot {\bf u}_a$.
Thus, the expansion of the fluid is equivalent to an entropy increase. Furthermore, we note that interference between plane waves leads to an increase of entropy, but it also renders the Madelung fluid compressible, allowing it to expand. Putting all of the pieces together, we find that a simple link exists between the interference of wave functions, compressibility of the Madelung fluid, and (Boltzmann) entropy production.

We illustrated how these concepts come to life in several specific examples. We underscored the difference between the Boltzmann and von Neumann entropies of the Madelung fluid (and derived simple new form for the latter).

The Madelung fluid expansion suggests an analogy with diffusive processes such as a free Joule expansion.
Ref. \cite{HC} related the imaginary part of the quantum velocity to the thermal fluctuations of the Madelung fluid.
Following the current analysis, this imaginary part can indeed be interpreted, as a diffusive drift velocity, since its flux is propositional to the minus sign of the density gradient (as in Fick's first law).
Furthermore, this analogy suggests that the diffusion coefficient is $\hbar/2m$, and for the simplest model of diffusion in ideal gas, the relation between the diffusion coefficient, the mean free path, and Brownian momentum provides the lower bound of Heisenberg's uncertainty relation.

The entropy production can be expressed as well in terms of the correlation expectation value between the advective and diffusive velocities. Moreover, for classical diffusive processes, it was shown that the gradient of the Bohm potential acts as a force to accelerate the diffusion. This appearance of the quantum potential in a classical mechanism may shed light on its role in the Schr\"{o}dinger equation. 
It is important to remember however, that in the quantum case the diffusion equation, governed by the Fick's second law, cannot be extracted from the Schr\"{o}dinger equation, i.e., $\der{\rho}{t} \neq (\hbar/2m) \nabla^2 \rho$, but rather $\der{\rho}{t} = -\nabla\cdot(\rho {\bf u}_a)$. Therefore, processes associated with density variation, such as entropy growth, are due to expansion by the hydrodynamical advective velocity and not by diffusive irreversible processes. \\

%As a last remark, we note that the Madelung fluid is a a Hamiltonian system and the fact that its entropy is not conserved is concerning. This may result from the fact that the Gibbs like entropy integral is calculated in space coordinates and not in the phase-space one.

\noindent {\bf Acknowledgements}\\
E.C. was supported in part by the Israel Science Foundation Grant No. 1311/14 and by ERC AdG NLST.

%
%...

\section*{Appendix}

Let our system be described by a continuous wavefunction $|\psi\rangle=\int dx \psi(x)|x\rangle$ (to simplify the notation we shall use in the appendix a 1-dimensional system, although the results are completely general). Then the pure density matrix would be:
\be
\hat{\rho}=|\Psi\rangle\langle\Psi|=\int\int dx dx' \Psi^*(x)\Psi(x')|x\rangle\langle x'|,
\ee
with entries $\Psi^*(x)\Psi(x')$. Therefore, in a basis where $\hat{\rho}$ is diagonal
\be
\begin{array}{lcl}
\hat{\rho} ln\hat{\rho}=\\
=\int\int\int\int dx dx' dy dy'  \Psi^*(x)\Psi(x') ln[\Psi^*(y)\Psi(y')]|x\rangle\langle x'|y\rangle\langle y'|=\\
=\int\int\int\int dx dx' dy dy'  \Psi^*(x)\Psi(x') ln[\Psi^*(y)\Psi(y')]\delta(x'-y)|x\rangle\langle y'|=\\
=\int\int\int dx dy dy'  \Psi^*(x)\Psi(y) ln[\Psi^*(y)\Psi(y')]|x\rangle\langle y'|=
\end{array}
\ee
To evaluate the von Neumann entropy (\ref{vne}) we shall now multiply the previous equation by $\langle z|$ from left, by $|z\rangle$ from right and then integrate over $z$ do find:
\be
\begin{array}{lcl}
Ent_{VN}=-\int\int\int\int dx dy dy' dz \Psi^*(x)\Psi(y)ln[\Psi^*(y)\Psi(y')]\langle z|x\rangle\langle y'|z\rangle=\\
=-\int\int\int\int dx dy dy' dz \Psi^*(x)\Psi(y)ln[\Psi^*(y)\Psi(y')] \delta(x-z)\delta(y'-z)=\\
=-\int\int dy dz \Psi^*(z)\Psi(y)ln[\Psi^*(y)\Psi(z)]=\\
=-\int\int dx dx' \Psi^*(x)\Psi(x')ln[\Psi^*(x')\Psi(x)],
\end{array}
\ee
where we only changed variables in the last line.

The von Neumann entropy of an isolated system is constant under unitary time evolution (this follows, e.g., from the cyclicity of the trace). We will show that explicitly:
\be
\begin{array}{lcl}
-\frac{\partial Ent_{VN}}{\partial t}=\int\int dx dx' \{\partial_t \Psi^*(x)\Psi(x')ln[\Psi^*(x')\Psi(x)]+\\
\Psi^*(x)\partial_t\Psi(x')ln[\Psi^*(x')\Psi(x)]+\\
\Psi^*(x)\Psi(x')\frac{\partial_t\Psi^*(x')}{\Psi^*(x')}+\\
\Psi^*(x)\Psi(x')\frac{\partial_t\Psi(x)}{\Psi(x)}\}.
\end{array}
\ee
We now use the Schr\"{o}dinger equation for evaluating all the time derivatives. Next, we differentiate twice by parts the first term. It cancels the last term while leaving an additional $-i \Psi^*(x)\Psi(x')\frac{(\partial_{x}\Psi(x))^2}{(\Psi(x))^2}$ (up to a multiplicative constant). Similarly, when differentiating twice by parts the second term, it cancels the third term, leaving an additional $i \Psi^*(x)\Psi(x')\frac{(\partial_{x'}\Psi^*(x'))^2}{(\Psi^*(x'))^2}$.
Therefore,
\be
\begin{array}{lcl}
-\frac{\partial Ent_{VN}}{\partial t}= i \int\int dx dx' \Psi^*(x)\Psi(x')[\frac{(\partial_{x'}\Psi^*(x'))^2}{(\Psi^*(x'))^2}-\frac{(\partial_{x}\Psi(x))^2}{(\Psi(x))^2}]\equiv Int.
\end{array}
\ee
The von Neumann entropy and its derivatives are real, hence $Int=Int^*$. However, if we perform a change of variables in $Int^*$ such that $x\rightarrow x'$ and $x' \rightarrow x$, we find out that $Int^*=-Int$. Therefore, $Int=0$ as required.

\end{document}